\documentclass[10pt,conference]{IEEEtran}
\IEEEoverridecommandlockouts

\usepackage{xcolor}
\usepackage{url}
\usepackage{xurl}
\usepackage[hidelinks]{hyperref}
\usepackage{amsfonts} 
\usepackage{algorithm}
\usepackage{algpseudocode}
\usepackage{amsmath}
\usepackage{array}
\usepackage{graphicx}

\usepackage{cite}
\usepackage{amssymb}
\usepackage{textcomp}
\usepackage{eso-pic}
\begin{document}


\title{\LARGE ARMOR-IMC: Adaptive Resource Mapping for Operational Robustness via Secure In-Memory Computing
}

\author{
\IEEEauthorblockN{
Muhtasim Alam Chowdhury\IEEEauthorrefmark{1},
Ramtin Zand\IEEEauthorrefmark{2},
Soheil Salehi\IEEEauthorrefmark{1}
}
\IEEEauthorblockA{
\IEEEauthorrefmark{2}Department of Computer Science and Engineering, University of South Carolina, Columbia, SC, USA\\
\IEEEauthorrefmark{1}Department of Electrical and Computer Engineering, University of Arizona, Tucson, AZ, USA\\
\{mmc7, ssalehi\}@arizona.edu\IEEEauthorrefmark{1}, ramtin@cse.sc.edu\IEEEauthorrefmark{2}
}
}



\maketitle
\AddToShipoutPictureFG*{%
  \AtPageLowerLeft{%
    \hspace{0.70in}%
    \raisebox{0.30in}{%
      \scriptsize 979-8-3195-0489-0/26/\$31.00~\copyright2026 IEEE%
    }%
  }%
}


\begin{abstract}
The massive data-movement overhead in traditional architectures has led to the adoption of In-Memory Computing (IMC) for energy-efficient Deep Neural Network (DNN) processing. By leveraging emerging devices like Spin-Orbit Torque Magnetic Tunnel Junctions (SOT-MTJs), IMC bypasses the “memory wall” and reduces leakage power inherent in traditional CMOS. However, this shift introduces dual hardware threats: manufacturing Process Variation (PV) degrades reliability and increases vulnerability to fault injection, while power Side-Channel Attacks (SCAs) compromise security.
Existing defenses address these threats in isolation. 
This work presents a post-training framework that simultaneously hardens analog IMC accelerators against both threats without retraining the model. 
Implemented in the IMAC-Sim simulator, our approach uses the proposed Variation Impact Score (VIS) to guide the mapping of Fault Observation Windows (FOWs) and introduces the Leakage Per Inference (LPI) metric to quantify input-dependent power variability under stochastic injection and the resulting reduction in effective signal-to-noise ratio.
Experiments show that PV-induced faults can degrade accuracy by over 50\%, while our method restores near-baseline accuracy and mitigates the threat of correlation-based power analysis attacks. 
\end{abstract}
\renewcommand\IEEEkeywordsname{Keywords}

\vspace{-0.2em}
\section{Introduction}
Deep Neural Networks (DNNs) are now integral to mission-critical systems, placing immense strain on conventional hardware. To overcome the resulting von Neumann bottleneck, the underlying hardware is shifting towards In-Memory Computing (IMC) architectures that leverage emerging devices like Spin-Orbit Torque Magnetic Tunnel Junction (SOT-MTJ) based Magnetic Random Access Memories (MRAMs) \cite{10319731}. However, this transition introduces significant hardware security and reliability challenges that threaten the integrity and confidentiality of AI workloads. The first major threat is reliability degradation caused by manufacturing Process Variation (PV). Our prior work demonstrated that subtle, adversarial, or unintentional variations in physical parameters, such as the SOT-MTJ device's oxide thickness ($t_{ox}$), can alter its resistive states, inducing systemic bit-flips in the MRAM weight arrays that significantly degrade the inference accuracy of the deployed DNN model \cite{6stune, 10528782}. The second major threat is physical Side-Channel Attacks (SCAs). It is well-established that DNN accelerators are vulnerable to attacks where an adversary with physical proximity measures the device's power consumption to infer secret data. Since the power consumed by a circuit is data-dependent, attackers can leverage statistical techniques such as Correlation Power Analysis (CPA) and Differential Power Analysis (DPA) to recover the secret parameters of the model, potentially compromising its Intellectual Property (IP) \cite{10.5555/3698900.3699091}. While some defenses exist, they typically address these reliability and security threats in isolation, leaving a critical gap where a system hardened against one threat remains vulnerable to the other.

\begin{figure}[!t]
    \centering
    \includegraphics[width=0.95\columnwidth]{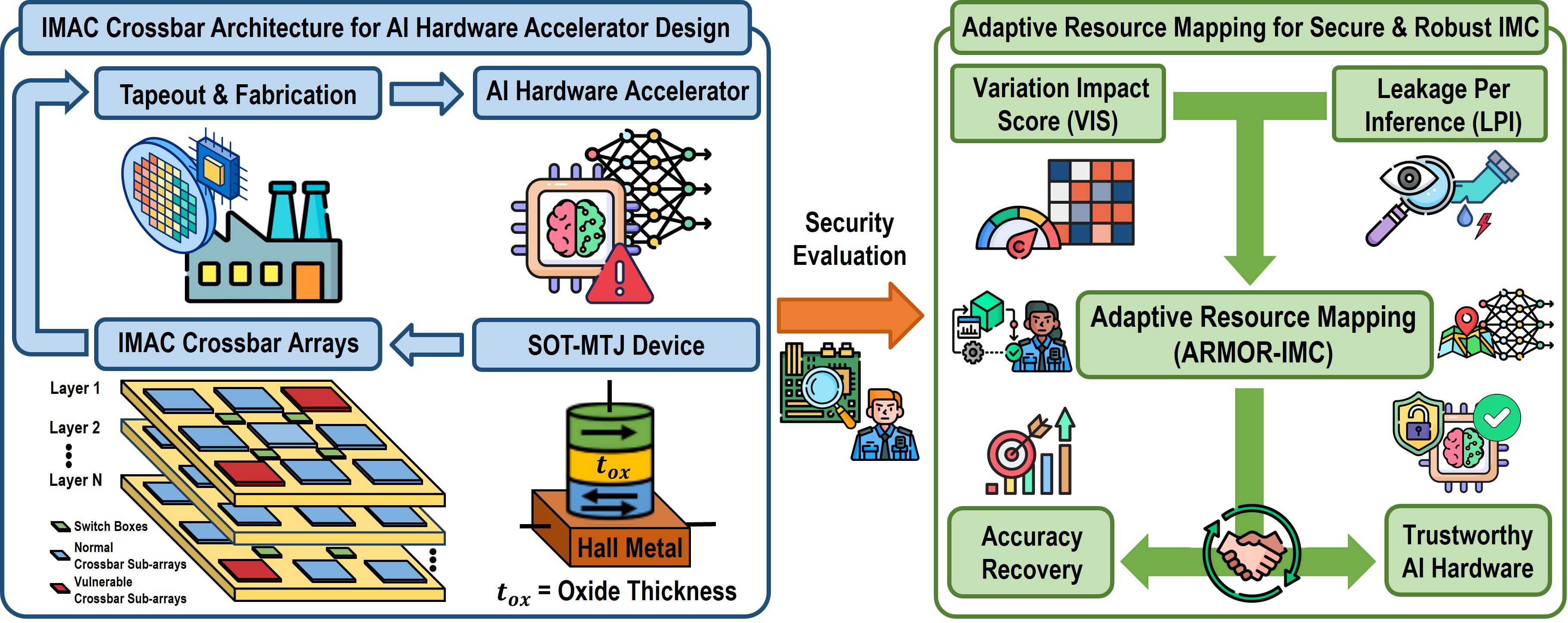}
        \vspace{-0.7em}
\caption{Overview of the ARMOR-IMC framework, illustrating the overall scope and highlighting the dual-metric analysis and adaptive mapping flow for improving both reliability and side-channel resilience in IMC accelerators.}
    \label{fig:overview}
    \vspace{-0.1em}
\end{figure}

To address this gap, we propose ARMOR-IMC, illustrated in Figure \ref{fig:overview}, a post-training framework that mitigates both PV-induced reliability issues and side-channel leakage vulnerabilities without requiring costly model retraining. Our entire methodology is implemented and validated within IMAC-Sim, a Python-based, circuit-level simulation framework that generates SPICE netlists of IMC circuits, enabling highly accurate evaluation of performance metrics, including power, latency, and the effects of interconnect parasitics \cite{7}. 
The core of our framework combines an architecture-level mapping strategy with a stochastic power injection mechanism, employing two complementary metrics defined over the FOW abstraction for reliability analysis and system-level power behavior for security analysis. The first metric, a Variation Impact Score (VIS), quantifies the reliability risk by running systematic fault injection campaigns to measure the accuracy degradation caused by random faults within each FOW. 
The second metric, Leakage Per Inference (LPI), quantifies the security risk by capturing input-dependent power variability.

\vspace{-0.1em}
\section{Background and Motivation} 
\label{sec:background}
Building upon the operational risks identified in mission-critical systems, we examine PV-induced bit-flip errors and data-dependent power leakage in analog IMC under the following threat model. We employ a gray-box adversarial model in which a supply-chain insider with limited visibility into the IMC design can manipulate fabrication processes to induce defects, while a non-invasive side-channel adversary monitors aggregate power consumption during inference.
\vspace{-0.3em}
\subsection{In-Memory Computing (IMC) and Process Variation}
Within the broader field of IMC, architectures are typically classified as either digital or analog, with this work focusing on analog IMC that utilizes the physical device properties of non-volatile memories such as SOT-MRAM to execute large-scale matrix computations such as Matrix Multiplication (MatMul) by exploiting Ohm's and Kirchhoff's laws. Even when accounting for the overhead of required data converters, the reduction in core computational energy allows analog IMC to achieve a lower total energy per operation compared to digital IMCs \cite{1}. To model this, we utilize IMAC-Sim, which has gained attention for its ability to efficiently model MatMul using crossbar arrays of memristive architectures \cite{7}. 
Despite these advantages, SOT-MTJ devices remain highly susceptible to process variation at scaled technology nodes.
These manufacturing fluctuations manifest in critical physical parameters such as the free-layer dimensions and, most significantly, the oxide tunnel barrier thickness ($T_{ox}$), which directly impacts the stability of the device's Parallel ($R_P$) and Anti-Parallel ($R_{AP}$) resistive states~\cite{6stune, 10528782}. Simulations using physically grounded Verilog-AMS models demonstrate that $T_{ox}$ variations exert an exponential influence on device behavior. In particular, if $T_{ox}$ is reduced by less than 1 nm, the resulting read current through the device can exceed the critical switching threshold, triggering unintentional bit-flips during read operations~\cite{6stune}. 
\vspace{-0.3em}
\subsection{Power Side-Channel Leakage in IMC Accelerators}
Power SCA exploits the fundamental dependence between a circuit's instantaneous power consumption and the data being processed, a principle formalized by Kocher et al. through Differential Power Analysis (DPA) and Correlation Power Analysis (CPA)~\cite{Kocher2011IntroductionTD, 10.5555/3698900.3699091}. While early SCA studies primarily targeted cryptographic implementations, recent work has demonstrated that machine learning accelerators are similarly vulnerable, as data-dependent switching activity can leak information about internal computations and model parameters. In the context of IMC, this risk is amplified by the stationary mapping of NN weights and the deterministic execution patterns of analog and mixed-signal compute macros. 
\vspace{-0.2em}
\section{Proposed Methodology: ARMOR-IMC}
\label{sec:methodology}
This section introduces the ARMOR-IMC framework.

\begin{figure}[!t]
    \centering
    \includegraphics[width=0.7\columnwidth]{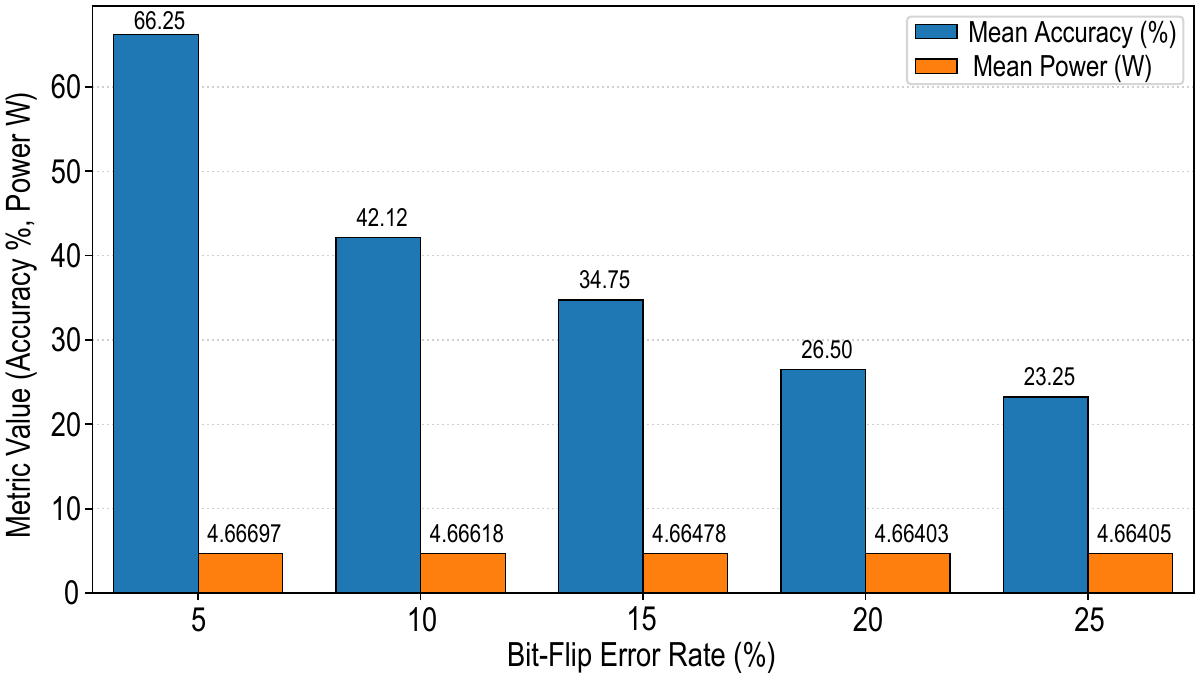}
       \vspace{-1em}
 \caption{Mean inference accuracy and mean power consumption across varying uniform bit-flip error rates (5\%–25\%).}
    \label{fig:accuracy_power_mean}
    \vspace{-0.2em}
\end{figure}
\vspace{-0.3em}
\subsection{VIS-Based Reliability Analysis and Remapping}
Our reliability analysis is based on introducing random bit-flips into the memristive crossbar arrays in order to identify vulnerable regions.  
These PV effects manifest as unintended, statistically distributed resistance shifts across SOT-MTJ devices, directly inducing bit-flip errors in stored NN weights~\cite{Huff_2021}.
The fault injection procedure is implemented using IMAC-Sim~\cite{7}, where software-trained NN weights are mapped onto resistive crossbar arrays using low-resistance (5 K$\Omega$) and high-resistance (15 K$\Omega$) states. 
PV-induced bit-flip errors are emulated by directly altering these resistance values, with 100-run Monte Carlo simulations per configuration. As shown in Figure~\ref{fig:accuracy_power_mean}, mean inference accuracy progressively degrades from 66.25\% at 5\% bit-flips to 23.25\% at 25\% fault rates, while mean power consumption remains stable. 

To enable fine-grained fault sensitivity analysis, IMAC-Sim partitions each NN layer into smaller sub-arrays that realistically model interconnect parasitics and physical layout constraints. In this work, we evaluate a three-layer DNN trained on the MNIST dataset with a $400\times120\times84\times10$ architecture, corresponding to $48{,}000$, $10{,}080$, and $840$ weights in layers 1, 2, and 3, respectively. Among supported sub-array sizes (32$\times$32, 64$\times$64, 128$\times$128, and 256$\times$256), the 32$\times$32 configuration provides the most reliable performance and is therefore adopted throughout this study. This partitioning yields fixed-size tiles referred to as \textit{Fault Observation Windows (FOWs)}, which represent the fundamental analytical units for localized fault analysis. The number of horizontal and vertical partitions for each layer is computed as:
\vspace{-0.5em}
\begin{equation}
H_p=\left\lceil\frac{N_{out}}{X_{row}}\right\rceil,\quad
V_p=\left\lceil\frac{N_{in}+1}{X_{col}}\right\rceil
\vspace{-0.5em}
\end{equation}
where $N_{in}$ and $N_{out}$ denote the number of input and output neurons, and $X_{row}$ and $X_{col}$ correspond to the FOW dimensions (32$\times$32). For example, the first layer is segmented into $4\times13=52$ FOWs, which are indexed for targeted fault injection and analysis.
Using this FOW-level abstraction, we define the VIS to quantify the sensitivity to localized faults within each layer. VIS is computed by individually injecting high bit-flip rates (50\%-90\%) into each FOW while keeping all other regions fault-free, followed by inference evaluation over 100 test images in each run. These localized PV experiments isolate fault-sensitive crossbar regions by aggregating VIS values into a scoreboard that ranks FOWs from most to least vulnerable based on their impact on inference accuracy.

\textbf{Dynamic Weight Mapping Strategy.}
To improve robustness against PV-induced faults, a dynamic weight mapping strategy is employed to reallocate critical weights from vulnerable regions of the crossbar to more reliable locations. 
The procedure operates at the granularity of FOWs and is guided by the computed VIS profiles. For a given weight matrix $W \in \mathbb{R}^{C \times R}$ partitioned into $N$ FOWs $\{\mathcal{S}_1,\dots,\mathcal{S}_N\}$, the most fault-affected region is identified by computing the Mean Absolute Error (MAE) between the fault-perturbed matrix $W$ and a clean reference matrix $W^{\mathrm{ref}}$ for each FOW:
\vspace{-0.2em}
\begin{equation}
\label{eq:mae}
\mathrm{MAE}_b = \frac{1}{|\mathcal{S}_b|}\sum_{(i,j) \in \mathcal{S}_b} \bigl|W_{ij} - W^{\mathrm{ref}}_{ij}\bigr|,
\end{equation}
and selecting the FOW with the maximum deviation as
\vspace{-0.5em}
\begin{equation}
\label{eq:bstar}
b^{\star} = \arg\max_b \mathrm{MAE}_b.
\vspace{-0.5em}
\end{equation}
A resilient partner FOW $k^{\star}$ is then selected from geometry-compatible candidates within the same layer, prioritized by highest fault-stressed inference accuracy derived from VIS. In the case of multiple candidates with identical fault-stress accuracy, ties are resolved by selecting the FOW with the lowest average power consumption. The contents of $\mathcal{S}_{b^{\star}}$ and $\mathcal{S}_{k^{\star}}$ are subsequently swapped, effectively relocating high-importance weights to more reliable physical regions without modifying the overall crossbar structure or operation count. This mapping is performed as a pre-inference step prior to SPICE netlist synthesis in IMAC-Sim.


\vspace{-0.3em}
\subsection{Leakage Per Inference (LPI) Metric} 

To establish exploitable power side-channel leakage without performing explicit CPA, we characterize baseline input-dependent power under a fixed, fault-free IMC mapping and test whether input data and power consumption exhibit a stable, learnable relationship. LPI is defined as the standard deviation of average power consumption measured across input inferences and is used as a first-order Leakage Power Analysis (LPA) indicator of input-dependent power variability.
Table~\ref{tab:lpi_baseline} reports the measured power statistics for each MNIST digit class, computed over 400 inference samples under identical hardware conditions. The results exhibit systematic variation in mean power across digits, with values ranging from 4.60~W (digit~0) to 4.68~W (digit~6). 
The inter-class separation exceeds the within-class standard deviation, indicating sufficient input-dependent leakage to motivate correlation-based SCA analysis.
Although architectural remapping is applied to mitigate this leakage, LPI remains unchanged ($\approx 0.059$~W) across Layer 1 swap configurations, indicating that static remapping alone is insufficient to suppress input-dependent leakage.


\textbf{Deterministic and Stochastic Noise Injection.}
Before stochastic obfuscation, we evaluate deterministic power balancing as a baseline active countermeasure. For each input, a compensatory auxiliary current sink adds the required power offset $P_{\text{aux}}(i)$ to equalize total inference power, thereby removing first-order input-dependent variation. However, this requires oracle knowledge of each input's exact power deficit and creates a static signature vulnerable to averaging attacks. 
Deterministic balancing suppresses first-order LPI but leaves a repeatable signature, motivating stochastic randomization.
To overcome the limitations of deterministic equalization, stochastic noise injection is introduced in which the auxiliary current sink is randomized for each execution. Prior to deployment, the auxiliary injected-power bounds are selected to span the observed baseline power variation while exceeding the measurement noise floor. During operation, for each input image $i$ in execution run $t$, an auxiliary power component $P_{\mathrm{aux}}(i,t)$ is sampled from a uniform distribution,
\vspace{-0.5em}
\begin{equation}
P_{\text{aux}}(i,t) \sim \mathcal{U}(\Delta P_{\min}, \Delta P_{ \max}),
\vspace{-0.5em}
\end{equation}
where $\Delta P_{\min}$ and $\Delta P_{\max}$ denote the lower and upper bounds of the auxiliary injected-power range, not bounds on total inference power. The sampled auxiliary power is realized through a time-gated resistive load connected to the supply rail and activated exclusively during the corresponding inference window. The required resistance is computed as
\vspace{-0.5em}
\begin{equation}
R_{\text{aux}}(i,t) = \frac{V_{DD}^2}{P_{\text{aux}}(i,t)},
\vspace{-0.5em}
\end{equation}
and the total observed power is given by
\vspace{-0.5em}
\begin{equation}
P_{\text{obs}}(i,t) = P_{\text{base}}(i,t) + P_{\text{aux}}(i,t)
\vspace{-0.5em}
\end{equation}
Since the auxiliary power component is independently resampled for each execution, the resulting power side-channel is transformed into a non-stationary stochastic process, intentionally preventing the formation of stable and repeatable power signatures exploitable by correlation-based SCAs.
\begin{table}[t]
\centering
\setlength{\tabcolsep}{3pt}
\caption{Baseline input-dependent power characteristics for MNIST digit classes under fixed fault-free mapping.}
\label{tab:lpi_baseline}
    \vspace{-0.8em}
\resizebox{\columnwidth}{!}{  
\begin{tabular}{c|c|c||c|c|c}
\hline
\textbf{Digit} & \textbf{Mean Power (W)} & \textbf{Std (W)} &
\textbf{Digit} & \textbf{Mean Power (W)} & \textbf{Std (W)} \\
\hline
0 & 4.6046 & 0.044 & 5 & 4.6043 & 0.064 \\
1 & 4.6236 & 0.065 & 6 & 4.6784 & 0.041 \\
2 & 4.6735 & 0.055 & 7 & 4.6441 & 0.072 \\
3 & 4.6477 & 0.061 & 8 & 4.6732 & 0.060 \\
4 & 4.6394 & 0.082 & 9 & 4.6221 & 0.065 \\
\hline
\end{tabular}
}
    \vspace{-0.1em}
\end{table}

\vspace{-0.1em}
\section{Experimental Results and Discussion}
\label{sec:results}
This section evaluates the ARMOR-IMC framework under fault-injection and power-leakage scenarios.
\begin{figure}[!t]
    \centering
    \includegraphics[width=\linewidth]{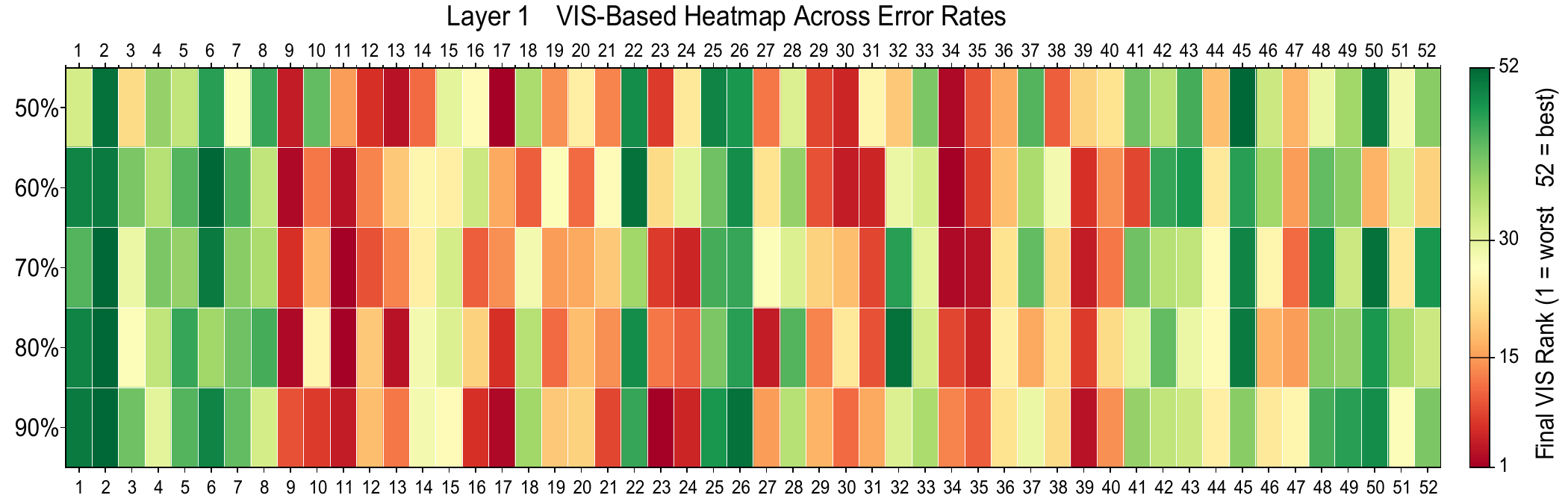}

    \vspace{1pt}

    \includegraphics[width=\linewidth]{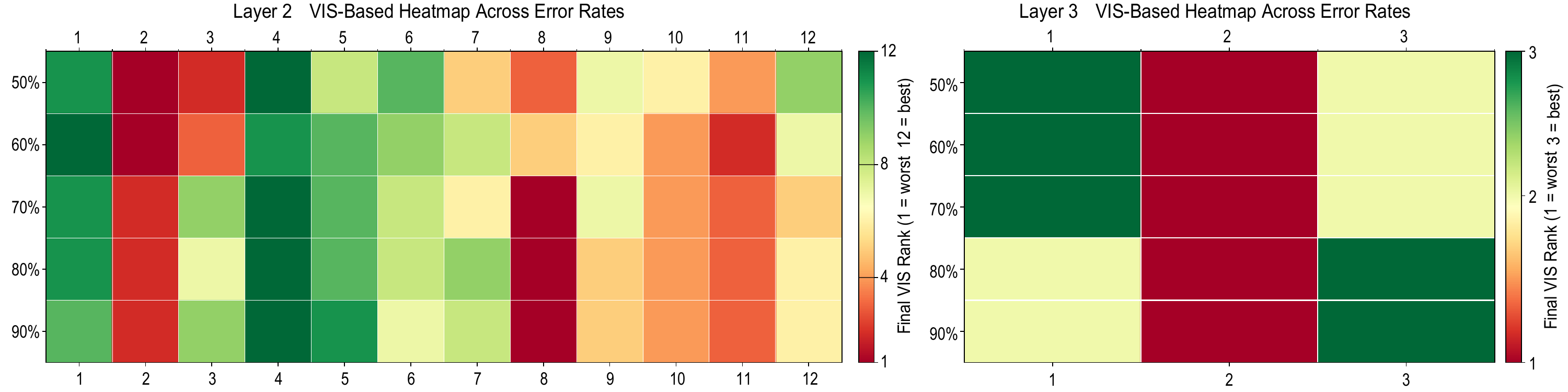}

        \vspace{-0.5em}
\caption{VIS heatmaps for all 3 layers after localized fault injection (50-90\%). 
    Top: Layer 1. Bottom: Layers 2 (Left) and~3 (Right). 
    Colors encode Final VIS error rank (red: high, yellow: moderate, green: low). }
    \label{fig:vis_all_layers_vertical}
    \vspace{-0.1em}
\end{figure}

\vspace{-0.3em}
\subsection{Reliability Preservation via VIS-Guided Remapping}
\begin{figure}[!t]
  \centering
  \includegraphics[width=\linewidth]{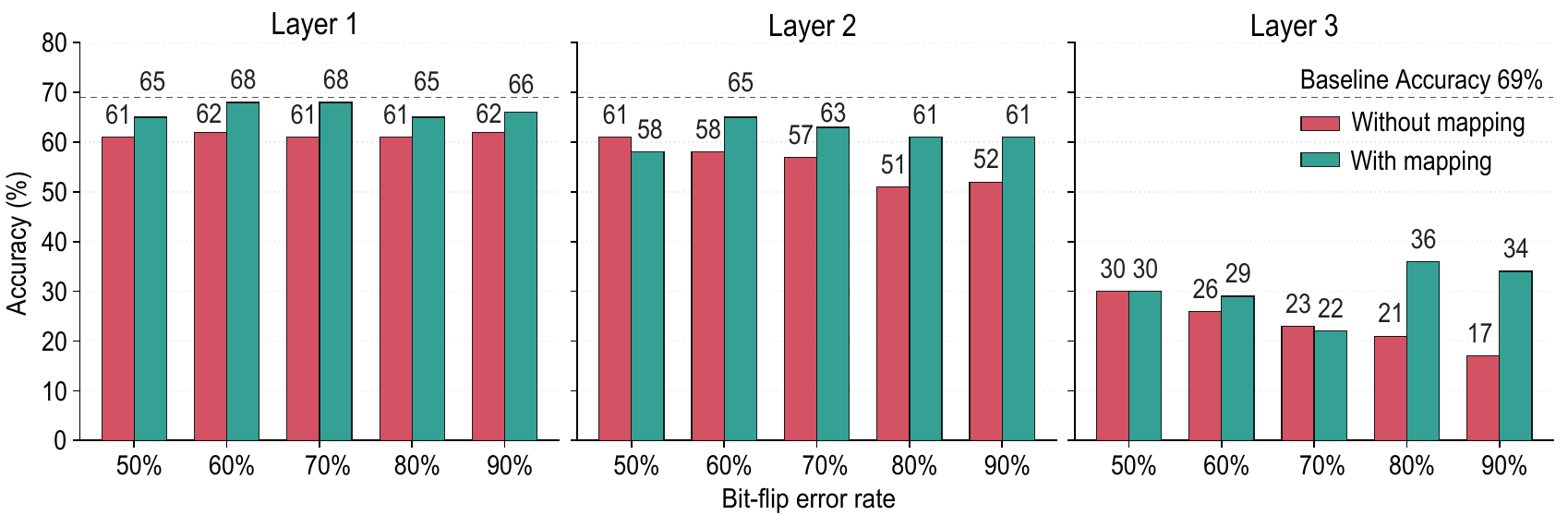}
      \vspace{-2em}
\caption{Accuracy under localized bit-flip faults with and without VIS-guided mapping across Layers 1-3 (50-90\% error rates), with Layer 3 showing the most pronounced recovery at higher fault-rates.}
  \label{fig:remap_accuracy_layers}
    \vspace{-0.2 em}
\end{figure}

The localized fault sensitivity of the SOT-MRAM crossbars was evaluated by applying bit-flip injection rates between 50\% and 90\% at the FOW granularity across all layers. As a reference, the error-free baseline achieves 69\% inference accuracy with 4.7 W average power consumption. As illustrated in the VIS heatmaps in Figure~\ref{fig:vis_all_layers_vertical}, the impact of these faults is highly non-uniform across layers. Layer~1 with 52 FOWs exhibits inherent resilience due to its large fan-in redundancy, maintaining approximately 61\% accuracy even at high fault rates, corresponding to a modest 7 to 8\% loss relative to the baseline. In contrast, Layer~3, consisting of only 3 FOWs, experiences severe degradation, with accuracy decreasing from 30\% at 50\% injection to 17\% at 90\% injection, representing a loss exceeding 50\% relative to the error-free case. 
Figure~\ref{fig:remap_accuracy_layers} demonstrates the efficacy of VIS-guided remapping in restoring this lost accuracy. The most significant recovery occurs in Layer~3, where remapping nearly doubles the accuracy at high fault-rates,  improving from 17\% to 34\% at 90\% injection by relocating the few critical defective regions to the limited set of available reliable FOWs. Layer~2 also shows strong recovery at high fault-rates; however, a minor degradation is observed at 50\% injection, where accuracy decreases from 61\% to 58\%. This anomaly arises because the faulty region corresponds to a $32\times20$ edge tile, and the strict constraint of swapping only between dimension-compatible FOWs restricts the candidate pool, forcing a suboptimal FOW partner selection.


\vspace{-0.3em}
\subsection{Input-Dependent Power Leakage Mitigation}
While the VIS-guided remapping effectively restores accuracy, baseline analysis confirmed that it leaves the fundamental power leakage signal intact. To mitigate this residual vulnerability, the deterministic power balancing successfully compresses the input-dependent power variation across digit classes, driving the LPI metric to very low values in both Layers~1 and~2 (LPI reduction $>96\%$). 
In both layers, average power values converge tightly around the baseline reference maximum, suppressing first-order amplitude differences and increasing the difficulty of naive threshold-based classification.
However, the static compensation rule renders power signatures stationary and predictable across runs, leaving the defense vulnerable to profiling and averaging attacks.

\begin{figure}[!t]
\centering
\includegraphics[width=0.65\columnwidth]{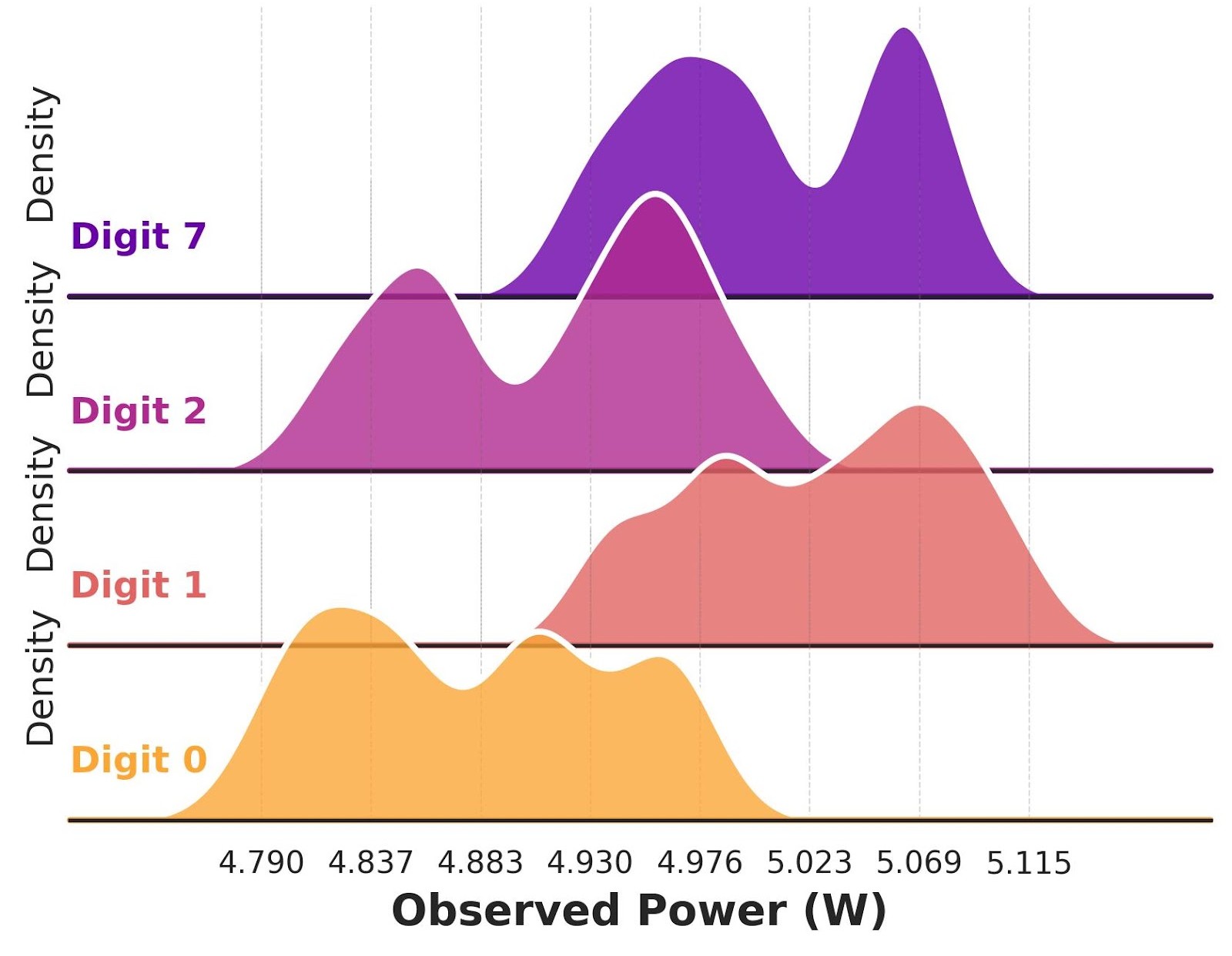}
    \vspace{-1em}
\caption{Ridge plot of input-dependent power distributions after stochastic power injection for selected digits. }
\label{fig:lpi_stochastic_overlap}
    \vspace{-0.2em}
\end{figure}
By injecting randomized noise sampled from a bounded distribution in the range $[0.2, 0.4]$~W during each inference window, we observe a significant increase in the variance of the observed power consumption, effectively masking the underlying input-dependent leakage. Analysis of the experimental data for Layer~1 reveals that the power distributions for different digit classes are no longer distinct “bands” but instead exhibit substantial overlap, as illustrated in Figure~\ref{fig:lpi_stochastic_overlap}. For instance, the power ranges for Digit~0 ($4.79-4.97$~W) and Digit~1 ($4.93-5.12$~W) overlap by approximately $34$~mW. Crucially, the total injected noise bandwidth ($\sim180$~mW) explicitly exceeds the baseline inter-class leakage signal ($\sim140$~mW). This overlapping behavior reduces the effective SNR, as the power signature of any input is masked by injected noise. 
Similar efficacy is observed in Layer~2 (overlap $\approx 53$~mW), indicating that stochastic injection reduces input-power correlation and increases the trace complexity of correlation-based SCAs. Consequently, LPI reflects the injected algorithmic noise magnitude, supporting effective SNR reduction below unity. Although this moving-target defense incurs a 6.5\% power overhead ($\sim$0.30~W), intermittent operation with randomized activation intervals keeps the leakage profile non-static and difficult to learn, providing increased resilience against simple and differential power analysis attacks.
\vspace{-0.1em}
\section{Conclusion}
\label{sec:conclusion}
This work presents ARMOR-IMC, a post-training framework for enhancing both reliability and security in analog IMC accelerators. The VIS metric enables fine-grained identification and mitigation of PV-induced fault sensitivity, while LPI captures input-dependent power variability and indicates side-channel leakage risk. The results demonstrate that VIS-guided mapping effectively restores inference accuracy under severe fault conditions and that stochastic auxiliary power injection renders power signatures nonstationary and harder to learn. 

\vspace{-0.1em}
\bibliographystyle{IEEEtran}
\bibliography{IEEEabrv,GOMACTech_LaTeX}
\end{document}